# A Similarity Transformed Second-order Approximate Coupled Cluster Method for the Excited States: Theory, Implementation, and Benchmark

Soumi Haldar and Achintya Kumar Dutta*

*Department of Chemistry, Indian Institute of Technology Bombay, Powai, Mumbai 400076, India*

**Abstract:**

We present a novel and cost-effective approach of using a second similarity transformation of the Hamiltonian to include the missing higher-order terms in the second-order approximate coupled cluster singles and doubles (CC2) model. The performance of the newly developed ST-CC2 model has been investigated for the calculation of excitation energies of valence, Rydberg, and charge-transfer excited states. The method shows significant improvement in the excitation energies of Rydberg and charge-transfer excited states as compared to the conventional CC2 method while retaining the good performance of the latter for the valence excited state. The method retains the charge-transfer separability of the CT excited states, which is a significant advantage over the traditional CC2 method. An MBPT2 variant of the new method is also proposed.

*achintya@chem.iitb.ac.in

## Introduction:

Accurate simulation of excitation energy is one of the most active areas of research in quantum chemistry. Among the various single-reference methods available for the calculation of excitation energy, the coupled-cluster method[1] has emerged as one of the most popular and systematically improvable ones. The ground state coupled-cluster method can be extended to excited states using the equation of motion (EOM) approach[2–6]. It leads to an identical final expression for the excitation energy as that of the linear response coupled cluster approach[7], although the two are derived from different theoretical viewpoints. The two approaches differ in the definition of transition property. The symmetry-adapted cluster CI method of Nakatsuji and co-workers[8] is also closely related to the former two approaches. The equation of motion coupled-cluster (EOM-CC) method is generally used in singles and doubles approximation (EOM-CCSD). It scales as O ($N^6$) power of the basis set and has a storage requirement similar to the ground state coupled cluster method, which restricts its use beyond small molecules. Therefore, the development of lower scaling approximations to EOM-CCSD is an active field of research[9]. One of the most successful approximations is the second-order approximate coupled cluster method (CC2) proposed by Jørgensen and co-workers[10]. Subsequently, a plethora of theoretical development within the framework of CC2 method has been described in the literature[11–18]. It includes efficient implementation based on density fitting[12,14,15,17] and semi-numerical approximation[19], local and natural orbital-based lower scaling approximation[10,11,13,14,18,20], calculation of first and second-order properties[21–24], solvation corrections[25], and analytic gradients[26]. Efficient CC2 implementations are available in many free and commercial software, which has made it the de facto standard for the correlated wave-function based excitation energy calculations on large molecules. Although the CC2 method gives excellent performance for the valence type of excited states, its performance deteriorates considerably for the Rydberg and charge-transfer states. Tajti and Szalay performed[27] a detailed analysis of the second-order approximate coupled cluster methods and concluded that the superior performance of the CC2 method for valence excited states is due to some fortuitous error cancellation between certain terms of the ground and excited states. However, such an error cancelation, unfortunately, does not take place for the Rydberg states. Tajti and Szalay[27] have advocated the explicit inclusion of certain missing terms in CC2 to restore back the balance for valence, and Rydberg states. However, such an attempt will increase the computational cost to O($N^6$) power of the basis set and will greatly reduce the computational advantages of the CC2 method.

In this paper, we propose an alternative approach of using a second similarity transformation of the Hamiltonian to include the missing higher-order terms in CC2 method without much increase in the computation cost.

## Theory:

**Canonical EOM-CC method:**

In the canonical EE-EOM-CCSD method[3], the target electronic excited state is obtained by the diagonalization of the coupled cluster (CC) similarity transformed Hamiltonian

$$\bar{H} = e^{-\hat{T}}\hat{H}e^{\hat{T}} \tag{1}$$

in a space spanned by 1 particle-1 hole and 2 particle-2 hole configurations. The cluster operator $\hat{T}$ is formed out of the ground state CC amplitudes

$$\hat{T} = \hat{T}_1 + \hat{T}_2 + \hat{T}_3 + .....\hat{T}_N \tag{2}$$

with

$$\hat{T}_m = \frac{1}{(m!)^2}\sum_{\substack{ij...\\ab...}} t_{ij...}^{ab...} a_a^\dagger a_b^\dagger ...a_j a_i \tag{3}$$

where the indices $i, j...$ refer to the occupied orbitals while the indices $a,b...$ refer to the virtual ones. The $t$ amplitudes in the singles and doubles (CCSD) approximation are obtained by the iterative solution of the following non-linear equations

$$\langle \Psi_i^a | \bar{H} | \Psi_0 \rangle = 0 \tag{4}$$

$$\langle \Psi_{ij}^{ab} | \bar{H} | \Psi_0 \rangle = 0 \tag{5}$$

The similarity transformed Hamiltonian has an approximate block structure

$$\begin{pmatrix} E_{CC} & \bar{H}_{0S} & \bar{H}_{0D} & 0 \\ 0 & \bar{H}_{SS} & \bar{H}_{SD} & \bar{H}_{ST} \\ 0 & \bar{H}_{DS} & \bar{H}_{DD} & \bar{H}_{DT} \\ \sim & \sim & \bar{H}_{TD} & \bar{H}_{TT} \end{pmatrix} \tag{6}$$

Where ~ indicates very small matrix elements introduced by three and higher body elementary operators.

The EOM-CCSD equation for the k[th] target state is given by

$$\bar{H}\hat{R}_k |\Psi_0\rangle = E_k \hat{R}_k |\Psi_0\rangle \tag{7}$$

where $\hat{R}_k$ is a linear excitation operator and $E_k$ is the energy of the k[th] target state.

The $\bar{H}$ being non-hermitian, there exists a left eigenvalue equation

$$\langle \Psi_0 | \hat{L}_k \bar{H} = \langle \Psi_0 | \hat{L}_k E_k \tag{8}$$

The right and the left eigenvectors follow the biorthogonality relation

$$\left\langle \hat{L}_i \middle| \hat{R}_j \right\rangle = \delta_{ij} \tag{9}$$

The energy difference of the ground and k[th] target state $(\omega_k = E_k - E_0)$ can be directly obtained from the commutator form of the equation (6).

$$\left[\bar{H}, \hat{R}_k\right] \left| \Psi_0 \right\rangle = \omega_k \hat{R}_k \left| \Psi_0 \right\rangle \tag{10}$$

The explicit expression of the $\hat{R}_k$ operator in the CCSD approximation for the excitation energy case is given by,

$$\hat{R}_k^{EE} = r_0 + \sum_{i,a} r_a^i a^\dagger i + \sum_{i>j,a>b} r_{ab}^{ij} a^\dagger i b^\dagger j \tag{11}$$

Equation (10) is solved by the modified Davidson iterative diagonalization method[28], which involves the contraction of the similarity transformed Hamiltonian with suitably chosen guess vectors to generate the so-called sigma vectors. The EOM part scales as O($N^6$) power of the basis set for the EE problem.

**STEOM-CCSD:**

Nooijen and co-workers[29] introduced a second similarity transformation of the Hamiltonian

$$\hat{G} = \{e^{\hat{S}}\}^{-1} \bar{H} \{e^{\hat{S}}\} = \{e^{\hat{S}}\}^{-1} e^{-\hat{T}} \hat{H} e^{\hat{T}} \{e^{\hat{S}}\} \tag{12}$$

where the operators $\hat{S}$ are normal ordered with respect to the reference function $\Psi_0$ and are defined as

$$\hat{S} = \hat{S}^{IP} + \hat{S}^{EA} \tag{13}$$

Within the singles and doubles approximation, the $\hat{S}^{IP}$ and $\hat{S}^{EA}$ are represented as

$$\hat{S}^{IP} = S_m^{i'} \{m^\dagger i'\} + \frac{1}{2} S_{mb}^{ij} \{m^\dagger i b^\dagger j\} \tag{14}$$

$$\hat{S}^{EA} = S_{a'}^{e} \{a'^\dagger e\} + \frac{1}{2} S_{ab}^{ej} \{a^\dagger e b^\dagger j\} \tag{15}$$

The index $m$ and $e$ denote active orbitals, which are basically subsets of occupied and virtual orbitals, respectively. The inactive occupied and virtual orbitals are denoted using primed labels ($i', a'$ ). The concept of active space in STEOM-CC is quite different from the active space in the complete active space (CAS) based method. The active space in STEOM-CC has nothing to do with the non-dynamic correlation and can be interpreted as the dynamic correlation of quasi-particles. The $\hat{S}$ amplitudes are chosen in such a way that the dominant coupling in the singles to doubles block of the matrix elements of $\hat{G}$ vanishes and one can get the excited state dominated by single excitation by diagonalizing $\hat{G}$ only in the CIS space. The $S$ amplitudes are obtained by setting the following blocks of the matrix elements of $\hat{G}$ to zero

$$g^{m}_{i'} = g^{a'}_{e} = 0 \tag{16}$$

$$g^{mb}_{ij} = g^{ab}_{ej} = 0 \tag{17}$$

It is the same as solving the effective Hamiltonian formulation of the Fock space multi-reference coupled cluster method[30,31] for (1,0) and (0,1) sectors. Alternatively, one can extract the $S$ amplitudes from the converged IP- and EA-EOM-CCSD eigenvectors, which is identical to solving the (0,1) and (1,0) sectors of Fock-space using the intermediate Hamiltonian Formulation of Meissner[32–35].
The matrix elements of the second similarity transformed Hamiltonian has the following structure

$$\begin{pmatrix} E_{CC} & G_{0S} & G_{0D} & 0 \\ 0 & G_{SS} & G_{SD} & G_{ST} \\ 0 & \simeq & G_{DD} & G_{DT} \\ \simeq & \simeq & \simeq & G_{TD} \end{pmatrix} \tag{18}$$

If one neglects the small residual elements $\simeq$ in the equation (18), one can obtain size-intensive excitation energies by diagonalizing $\hat{G}$ only in the space of single excitations.

Following ref 35, one can rewrite equation (12) as

$$\hat{G} = \{e^{\hat{S}_1}\}^{-1}\{e^{\hat{S}_2}\}^{-1}\bar{H}\{e^{\hat{S}_2}\}\{e^{\hat{S}_1}\} = \{e^{\hat{S}_1}\}^{-1}\hat{G}_2\{e^{\hat{S}_1}\} \tag{19}$$

Nooijen and co-workers[29] have demonstrated that diagonalization of $\hat{G}$ and $\hat{G}_2$ leads to identical eigenvalues, and in the STEOM-CCSD method, one actually diagonalizes $\hat{G}_2$. For more details about the theory of STEOM-CCSD, the readers should consult the original implementation paper by Nooijen and co-workers[29,36]. Recent years have seen a renewed interest in the STEOM-CCSD method. Efficient implementations based on local and natural orbitals[37,38], automatic active space selection scheme[39], the extension to the open-shell systems[40], vibrational problem[41], and spin-orbit coupling calculation[42] with the framework of STEOM-CCSD method have been reported.

**The CC2 method:**

In the CC2 method the excitation energies were obtained by the diagonalization of the effective Hamiltonian

$$A=\begin{pmatrix} \langle \Psi_i^a | \left[ \bar{H}^{CC2}, a_c^{\dagger} a_k \right] | \Psi_0 \rangle & \langle \Psi_i^a | \left[ \bar{H}^{CC2}, a_c^{\dagger} a_k a_d^{\dagger} a_l \right] | \Psi_0 \rangle \\ \langle \Psi_{ij}^{ab} | \left[ e^{-\hat{T}_1^{CC2}} H e^{-\hat{T}_1^{CC2}}, a_c^{\dagger} a_k \right] | \Psi_0 \rangle & \langle \Psi_{ij}^{ab} | \left[ F, a_c^{\dagger} a_k a_d^{\dagger} a_l \right] | \Psi_0 \rangle \end{pmatrix} \tag{20}$$

where

$$\bar{H}^{CC2} = e^{-\hat{T}^{CC2}} \hat{H} e^{\hat{T}^{CC2}} \tag{21}$$

The coupled cluster amplitudes for CC2 is obtained from the following equations

$$\langle \Psi_i^a | \overline{H}^{CC2} | \Psi_0 \rangle = 0 \tag{22}$$

$$\langle \Psi_{ij}^{ab} | e^{-\hat{T}_1^{CC2}} \hat{H} e^{\hat{T}_1^{CC2}} + \left[ F, \hat{T}_2^{CC2} \right] | \Psi_0 \rangle = 0 \tag{23}$$

The matrix elements of $\bar{H}^{CC2}$ has the following structure

$$\begin{pmatrix} E_{CC2} & \bar{H}_{0S} & \bar{H}_{0D} \\ Z_1 & \langle \Psi_i^a | e^{-\hat{T}_2^{CC2}} e^{-\hat{T}_1^{CC2}} \hat{H} e^{\hat{T}_1^{CC2}} e^{\hat{T}_2^{CC2}} | \Psi_k^c \rangle & \langle \Psi_i^a | e^{-\hat{T}_2^{CC2}} e^{-\hat{T}_1^{CC2}} \hat{H} e^{\hat{T}_1^{CC2}} e^{\hat{T}_2^{CC2}} | \Psi_{kl}^{cd} \rangle \\ Z_2 & \langle \Psi_{ij}^{ab} | e^{-\hat{T}_1^{CC2}} \hat{H} e^{-\hat{T}_1^{CC2}} | \Psi_k^c \rangle & \langle \Psi_{ij}^{ab} | \hat{F} | \Psi_{kl}^{cd} \rangle \end{pmatrix} \tag{24}$$

The block structure of $\bar{H}^{CC2}$ differs from the standard CCSD $\bar{H}$ in the presence of Z elements. They give the residual non-zero $\bar{H}_{S0}$ and $\bar{H}_{D0}$ matrix elements arising due to the approximate solution to the CCSD equations. In the framework of EOM-CC theory, the size-intensive energies can be obtained from the diagonalization of $\bar{H}^{CC2}$ by resorting to a time-tested technique of neglecting Z. One can define a projected Hamiltonian as

$$\begin{pmatrix} E_{CC2} & \bar{H}_{0S} & \bar{H}_{0D} \\ 0 & \langle \Psi_i^a | e^{-\hat{T}_2^{CC2}} e^{-\hat{T}_1^{CC2}} \hat{H} e^{\hat{T}_1^{CC2}} e^{\hat{T}_2^{CC2}} | \Psi_k^c \rangle & \langle \Psi_i^a | e^{-\hat{T}_2^{CC2}} e^{-\hat{T}_1^{CC2}} \hat{H} e^{\hat{T}_1^{CC2}} e^{\hat{T}_2^{CC2}} | \Psi_{kl}^{cd} \rangle \\ 0 & \langle \Psi_{ij}^{ab} | e^{-\hat{T}_1^{CC2}} \hat{H} e^{-\hat{T}_1^{CC2}} | \Psi_k^c \rangle & \langle \Psi_{ij}^{ab} | \hat{F} | \Psi_{kl}^{cd} \rangle \end{pmatrix} \tag{25}$$

which is equivalent to the CC2 effective Hamiltonian in eq (20).

Tajti and Szalay[27] have shown that the missing terms of $\bar{H}_{DS}$ block in the CC2 method approximately cancel the missing elements of $\bar{H}_{DD}$ block in case of the valence excitation and this error cancellation leads to superior results. Unfortunately, such an error cancellation does not take place for Rydberg states. It was further shown[27] that even the lowest order inclusion of the missing terms in $\bar{H}_{DS}$ and $\bar{H}_{DD}$ can result in restoring the balance for valence and Rydberg states. However, a minimal order inclusion of the missing terms of $\bar{H}_{DD}$ block, beyond the Fock operator, leads to an increased computation cost.

We have used an alternative approach of dressing of the $\bar{H}$ with the $\hat{S}$ to bring in a minimal description of the missing coupling elements of the doubles-doubles block. One can compress the information contained in the double excitation operator of EOM-CCSD to the singles space by using an effective Hamiltonian formulation of EOM-CCSD. Meissner and Bartlett[43] have shown that the effect of doubles in the EOM-CCSD picture can be recovered by diagonalization of an appropriately dressed Hamiltonian in the singles space. A similar effect can be considered for STEOM-CCSD, where diagonalization of $\hat{G}_2$ will bring in the effect of higher-order excitations. It should be noted that the three-body terms in the transformed Hamiltonian $\hat{G}_2$ that is responsible for the remaining coupling between the singly and doubly excited determinants are neglected in the STEOM-CCSD picture. It means that certain interactions that couple the hole and particle in the doubles-doubles block of the EOM-CCSD are absent in STEOM-CCSD[29]. The hole is correlated separately, and the particle is correlated separately in the STEOM-CCSD picture. The interactions between the two are absent in the doubles-doubles block of the second similarity transformed matrix. The other blocks in STEOM-CCSD contain the full coupling. However, it should not be the problem in the present context as we are only interested in a minimal inclusion of the coupling of doubles-doubles block.

We define an approximate second similarity transformed Hamiltonian

$$\hat{G}_2^{CC2} = \{e^{\hat{S}_2}\}^{-1} \bar{H}_A^{CC2} \{e^{\hat{S}_2}\} \tag{26}$$

where

$$\bar{H}_A^{CC2} = \begin{pmatrix} E_{CC2} & \bar{H}_{0S} & \bar{H}_{0D} \\ 0 & e^{-\hat{T}_2^{CC2}} e^{-\hat{T}_1^{CC2}} \hat{H} e^{\hat{T}_1^{CC2}} e^{\hat{T}_2^{CC2}} & e^{-\hat{T}_2^{CC2}} e^{-\hat{T}_1^{CC2}} \hat{H} e^{\hat{T}_1^{CC2}} e^{\hat{T}_2^{CC2}} \\ 0 & e^{-\hat{T}_2^{CC2}} e^{-\hat{T}_1^{CC2}} \hat{H} e^{\hat{T}_1^{CC2}} e^{\hat{T}_2^{CC2}} & \hat{F} \end{pmatrix} \tag{27}$$

One can see that the missing $\hat{T}_2$ contributions to the double-singles block of the CC2 effective Hamiltonian are included in the $\bar{H}_A^{CC2}$. However, the missing terms containing bare-integrals and the $\hat{T}_1$ and $\hat{T}_2$ containing terms of the doubles-doubles block is still missing, and their effect will be introduced by the dressed elements in the $\hat{G}_2^{CC2}$. If one neglects the singles amplitudes in $\bar{H}_A^{CC2}$ and diagonalizes it within the space of singly and doubly excited configurations, it will lead to P-EOM-CCSD-MBPT(2) method[44,45].

The $S^{IP}$ and $S^{EA}$ amplitudes are obtained by diagonalizing $\bar{H}_A^{CC2}$ in the space of 1h,2h1p and 1p,2p1h configurations, respectively, and by putting intermediate normalization conditions on the converged eigenvectors. The diagonalization of $\hat{G}_2^{CC2}$ within the singles space will lead to size-intensive excitation energies, and we call the resulting method as ST-CC2. The exact programmable expressions are given in the Appendix. Figure 1 shows the steps involved in ST-CC2. However, the current implementation as presented in this work is far from an optimized version and scales as non-iterative O($N^6$) in the present form. One can obtain an iterative O($N^5$) scaling implementation by constructing the $\bar{H}_A^{CC2}$ intermediates on the fly. However, such an approach is not considered here.

If one approximates the ground state amplitudes in ST-CC2 by MP2 amplitudes, it will lead to P-STEOM-MBPT2 method, which is a similarity transformed analog of P-EOM-MBPT2 method. In P-STEOM-MBPT2 method, the IP and EA-EOM-CCSD steps of STEOM-CCSD are replaced by partitioned IP[46] and EA-EOM-MBPT2 methods[44]. The P-STEOM-MBPT2 method has the same formal scaling as that of the ST-CC2 method and is computationally advantageous over the non-iterative O($N^6$) scaling of the P-EOM-MBPT2 method. The ST-CC2 and P-STEOM-MBPT2 methods are implemented in a development version of ORCA[47].

## 3. Result and Discussion:

The original CC2 approximation and its newly developed similarity transformed versions are mainly applicable to the excited states dominated by single excitation. The singly excited states are broadly classified into three categories: valence, Rydberg, and charge transfer. These three kinds of excited states have significantly different physics, and one needs to individually benchmark the accuracy of the ST-CC2 method for these kinds of excited states. All the ST-CC2 and P-STEOM-MBPT2 calculations are performed with a development version of ORCA.

### 3.A Valence excited states:

Thiel's test set[48] is a popular benchmark set for gauging the accuracy of valence excited states dominated by single excitation. Table 1 presents the statistical analysis of the errors in the excitation energies for CC2, P-STEOM-MBPT2 and ST-CC2 method for singlet and triplet states with respect to benchmark CC3[49] values. The individual excitation energies for singlet and triplet states are presented in Tables S1 and S2, respectively. The CC3 values are taken from the work of Nooijen and co-workers[50], except for the nucleobases. The CC3 results for the nucleobases are taken from the work of Szalay and co-workers[51]. Following the work of Nooijen and co-workers, the excited states with more than 87 percent of singles characters are considered for the analysis. The CC2 triplet results have been taken from the work of Thiel and co-workers[48].

The ST-CC2 method shows an average absolute deviation (AAD) of 0.101 eV for singlet excited states, which is slightly higher than that observed for the CC2 method (0.085 eV). The maximum absolute deviation (MAD) is also higher at 0.376 eV in the ST-CC2 method. The error bar of the ST-CC2 method is slightly higher than the parent STEOM-CCSD, for which an AAD of 0.07 eV has been reported in the literature[38]. The spread of the error is also larger in ST-CC2 (0.125 eV) than that observed in STEOM-CCSD (0.09 eV). However, the errors in ST-CC2 are less than that of the P-STEOM-MBPT2 method, which shows MAD and AAD of 0.503 eV and 0.136 eV, respectively. The other statistical parameters also follow the same trends. The error in the PSTEOM-MBPT2 method shows a larger spread than the ST-CC2 method (see Figure 2(a)) for singlet states. It should be noted that the EOM-CCSD method shows a MAD and AAD of 0.39 eV and 0.20 eV, respectively, for the same excited states[38].

One can try to rationalize the trends observed in ST-CC2 results by having a comparison with the other approximate variant of STEOM-CCSD. One can construct the standard CCSD $\bar{H}$ (equation (6)) with CC2 amplitudes and solve the following STEOM equations with this $\bar{H}$. The remaining non-zero elements of $\bar{H}_{S0}$ and $\bar{H}_{D0}$ needs to be projected out. We call this method as extended ST-CC2 (Ext-ST-CC2). On the other hand, one can construct the $\bar{H}_A^{CC2}$ (equation (27)) using the CCSD amplitudes and solve subsequent STEOM-CCSD equations using this modified the $\bar{H}_A^{CC2}$. We call this method partitioned STEOM-CCSD (P-STEOM-CCSD). One can see that the Ext-ST-CC2 has a similar ground state to the ST-CC2, whereas the P-STEOM-CCSD has an excited state description similar to the ST-CC2. To demonstrate the behaviors of these approximate STEOM-CCSD methods, we have presented the excitation energies corresponding to the singlet excited state of S-tetrazene in Table 2. It is gratifying to note that the performance of ST-CC2 as compared to STEOM-CCSD does not significantly deteriorate for excited states having higher double excitation character. Both approaches show similar deviations from the CC3 reference values. One can see that the Ext-ST-CC2 method overestimates and the P-STEOM-CCSD underestimates the excitation energies as compared to the STEOM-CCSD values for all the excited states irrespective of their double excitation character. Therefore, the enhanced accuracy of the ST-CC2 method is due to the error cancellation introduced by the truncated ground state and excited state wave functions. Figure 3 presents the error distribution plot for the valence singlet excited states dominated by single excitations in Thiel's test set with respect to a STEOM-CCSD reference. It shows that the error cancelation in ST-CC2 as observed for S-tetrazene in Table 2 is a general trend. Table S3 presents the statistical analysis of errors in the Ext-ST-CC2 and P-STEOM-CCSD methods with respect to a CC3 reference. The individual values of the excitation energies obtained from these methods are given in Table S4. One can see that both Ext-ST-CC2 and P-STEOM-CCSD methods show a slightly larger error than the ST-CC2 method. We have also considered the partitioned STEOM-CCSD approximation, where partitioned EOM-CCSD approximation is only considered for the IP case (PIP-STEOM-CCSD) or the EA case (PEA-EOM-CCSD), and the results are presented in Table S4. One can see that the effect of the terms beyond the diagonal one in the doubles-doubles block of the underlying IP-EOM-CCSD and EA-EOM-CCSD method is generally small on the STEOM-CCSD excitation energy, and the effect of higher-order terms mostly appears as a constant shift. Especially, the truncation of the doubles-doubles block of the EA-EOM-CCSD has a negligible effect on the accuracy of the STEOM-CCSD method.

Trends observed in the excitation energies obtained from the CC2 method are quite different for the triplet states. The error in CC2 excitation energies increases significantly for the valence triplet states. The AAD for triplet states in CC2 is 0.155 eV, which is almost double of that observed for the singlet states. The MAD in CC2 also increases to 0.56 eV for the triplet state. The MAD and AAD of valence triplet states in the EOM-CCSD method have been reported to be 0.51 eV and 0.11 eV, respectively[38]. It shows that the error cancellation due to missing terms of doubles-singles and doubles-doubles block of CC2 is not general and only restricted to valence singlet states. The ST-CC2 method shows slightly superior performance for valence triplet excited states than the conventional CC2 with MAD and AAD of 0.396 eV and 0.138 eV, respectively. The statistical parameters are in the same range as observed for the STEOM-CCSD[38]. The performance of ST-CC2 is quite consistent for singlet and triplet states. However, the spread of error for triplet states in ST-CC2 is slightly larger than the standard CC2 (See Figure 2) for both singlet and triplet states. The PSTEOM-MBPT2 method shows slightly inferior results as compared to ST-CC2 both in terms of the magnitude of the error and their spread.

### 3.B The Rydberg excited states:

The standard CC2 method is known to give significantly inferior results for Rydberg excited states[27]. To investigate the performance of the new ST-CC2 method for the singlet and triplet Rydberg excited states, we have used the Waterloo test set[52] of Nooijen. Ahlrich's TZVP basis set[53] has been used along with additional 5s4p4d diffuse functions added to the ghost atom placed at the center of symmetry for both singlet and triplet calculations. The CC3 results are considered as the reference value for the singlet states, whereas EOM-CCSD(T) results have been used as the reference for the triplet states. All the CC3 and EOM-CCSD(T) results have been taken from reference [38,52]. The statistical analysis of the results for both the singlet and triplet excited states of the Rydberg test set is presented in Table 3. The individual excitation energies for singlet and triplet states are presented in Tables S5 and S6, respectively.

The Rydberg singlet states in CC2 are significantly underestimated with respect to the benchmark CC3 values. The MAD in the CC2 method for Rydberg singlet states is as high as 0.985 eV. The AAD in the CC2 method is 0.27 eV for the singlet states. The errors are much higher than that observed for the EOM-CCSD, which shows a MAD and AAD of 0.39 eV and 0.08 eV, respectively. The trend is consistent with the previously reported CC2 results for Rydberg singlet states[54]. The ST-CC2 method, on the other hand, gives a performance comparable to that for the valence state with MAD and AAD of 0.276 eV and 0.104 eV, respectively. Figure 4(a) shows that the spread in error for Rydberg singlet states in ST-CC2 is also much less than that of the CC2 method. The errors in the ST-CC2 results are similar to STEOM-CCSD[38] method, which shows a MAD and AAD of 0.279 eV and 0.071 eV, respectively. The P-STEOM-MBPT2 is, on the other hand, shows a slightly inferior performance than the ST-CC2 with MAD and AAD of 0.375 eV and 0.155 eV, respectively. The spread in errors is also slightly higher in P-STEOM-MBPT2. However, the errors are much less than the errors observed in CC2.

No CC2 results are available for the Rydberg triplet states. However, the performance of the ST-CC2 method for triplet Rydberg states is similar to the singlet state with MAD and AAD of 0.309 eV and 0.087 eV, respectively with respect to an EOM-CCSD(T) reference. The ST-CC2 results are also in good agreement with standard EOM-CCSD and STEOM-CCSD methods, both of which show[38] an AAD of 0.049 eV. The P-STEOM-MBPT2 method gives a slightly inferior performance compared to the ST-CC2 with MAD and AAD of 0.449 eV and 0.157 eV, respectively. The spread of the error is also more prominent in P-STEOM-MBPT2 than that in the ST-CC2 (see Figure 4(b)).

**3.C The charge transfer excited states:**

Electronic excitation, where the excited electron gets transferred from one fragment to another within the same system, gives rise to charge-transfer (CT) states. One of the strong points of the standard STEOM-CCSD method is charge transfer separability i.e., the excitation energy for a charge transfer excitation between two fragments separated at infinite distance is the same as the sum of the ionization potential of one fragment and the electron affinity of the other.

$$EE = IP + EA - \frac{e^2}{R}$$

Where EE, IP, and EA are the charge transfer excitation energy, ionization potential of the donor fragment, and electron affinity of the acceptor fragment. The term $\frac{e^2}{R}$ represents the long-range Coulomb interaction between the two charged fragments. To investigate the charge transfer separability behavior of the ST-CC2, we have chosen the Be-$C_2$ model system. The model system is built by placing the $C_2$ unit perpendicular to the axis connecting Be and the midpoint X of $C_2$ (See Figure 5). The Be–X distance is then varied, keeping all other geometrical parameters fixed, and the corresponding charge transfer parameters are reported in Table 4 of the manuscript. One can see that both ST-CC2 and P-STEOM-MBPT2 demonstrate correct asymptotic $R^{-1}$ behavior with increasing inter fragment distances. The CC2 method, on the other hand, shows a large separability error of 0.2641 eV at the internuclear separation of 10000Å. The lack of charge transfer separability in CC2 is similar to its parent EOM-CCSD method.

To see the implication of charge transfer separability on the accuracy of excitation energy for the charge-transfer excited states, we have considered the new benchmark set proposed by Szalay and co-workers[55]. The test set consists of a total of ten dimers having low energy CT states. The ST-CC2 and P-STEOM-MBPT2 results are calculated for the benchmark set in the cc-pVDZ basis set. The CC2 and the reference EOM-CCSDT-3 results used for all 14 CT states are taken from the ref[55]. Table 5 presents the statistical analysis of the results. The ST-CC2 methods show the best performance with MAD and AAD of 0.564 eV and 0.163 eV. The P-STEOM-MBPT2 shows slightly inferior performance with MAD and AAD of 0.633 eV and 0.187 eV, respectively. The spread of the error is also higher in the P-STEOM-MBPT2 methods (See Figure 6). The CC2 method, on the other hand, shows a much larger error with MAD and AAD of 0.76 eV and 0.363 eV, respectively. It shows that the lack of charge transfer separability also gets manifested into the larger error in charge-transfer states for CC2. Similar behavior was previously observed for the charge transfer states in the parent EOM-CCSD method[55], which shows an AAD of 0.296[23]. Although the ST-CC2 results for charge transfer states are much better than the standard CC2 method, the AAD is almost twice that observed for the STEOM-CCSD method (0.076) eV. Similar behavior is observed for the P-STEOM-MBPT2 method.

**4. Conclusions:**

We describe the theory and implementation of a second similarity transformed version of the second-order approximate coupled cluster singles and doubles method. The method retains the good performance of CC2 method for valence singlet and triplet excited state and greatly reduces the error for Rydberg and charge-transfer states. The results in the new ST-CC2 method closely follow the trends in the parent STEOM-CCSD method for the excited states dominated by single excitations and gives balanced performance for valence, Rydberg, and charge-transfer states. The excitation energies obtained from the ST-CC2 method demonstrate charge transfer separability, the lack of which leads to a larger error in the CC2 method. The MBPT2 variant of the ST-CC2 shows similar trends, but the errors are slightly more than the original ST-CC2. It shows that the terms beyond the diagonal elements in the doubles-doubles block of IP- and EA-EOM-CCSD have a minimal effect on the STEOM-CCSD energy. Therefore, even a minimal inclusion of such terms (maybe using a natural orbital basis) will lead to high accuracy in the resulting STEOM-CCSD method.

Work is in progress towards the development of a natural orbital-based implementation of STEOM-CCSD.

**Supporting Information:**

All the excitation energies for the singlet and triplet state of valence, Rydberg, and charge-transfer test set are provided in the supporting information.

**Acknowledgment**

The authors acknowledge the support from the IIT Bombay, IIT Bombay Seed Grant project, DST-Inspire Faculty Fellowship for financial support, DST SERB, IIT Bombay super computational facility, and C-DAC Supercomputing resources (PARAM Yuva-II, Param Bramha) for computational time.

Conflict of interest

The authors declare no competing financial interest.

**Appendix:**

Sigma equations for IP

$$\sigma^i = \underbrace{-\tilde{\tilde{F}}_i^l r^l + \tilde{F}_d^l \tilde{r}_d^{il}}_{F} - \underbrace{\hat{g}_{id}^{lm} \tilde{r}_d^{lm}}_{3i} .$$

$$\sigma_b^{ij} = \underbrace{F_d^b r_d^{ij} - F_j^l r_b^{il} - F_i^l r_b^{lj}}_{F} - \underbrace{\tilde{g}_{lb}^{ij} r^l}_{3i} .$$

Sigma equations for EA

$$\sigma_a = \underbrace{\tilde{\tilde{F}}_d^a r_d + \tilde{F}_d^l \tilde{r}_{ad}^l}_{F} + \underbrace{\hat{g}_{ed}^{la} \tilde{r}_{de}^l}_{3e}$$

$$\sigma_{ab}^j = \underbrace{F_d^a r_{db}^j + F_d^b r_{ad}^j - F_j^l r_{ab}^l}_{F} + \underbrace{\tilde{g}_{ba}^{jd} r_d}_{3e} .$$

The $\tilde{F}$ , $\tilde{\tilde{F}}$ , $\tilde{g}$ , $\hat{g}$ , $\tilde{K}$ , $\tilde{\tilde{K}}$ and $\tilde{J}$ are standard $\overline{H}$ intermediates and their expressions can be found in reference[56]. The expression for the ground state CC2 amplitudes can be found in ref 10.

The sigma vector for final ST-CC2 diagonalization

Triplet state

$$\sigma_i^a = \overline{\overline{g}}_a^b r_i^b - \overline{\overline{g}}_j^i r_j^a - \overline{\overline{g}}_{ab}^{ij}(J) r_j^b .$$

Singlet state

$$\sigma_i^a = \overline{\overline{g}}_a^b r_i^b - \overline{\overline{g}}_j^i r_j^a + 2\overline{\overline{g}}\left(K\right)_{ab}^{ij} r_j^b - \overline{\overline{g}}_{ab}^{ij}(J) r_j^b .$$

The expressions for the G intermediates

$$\tilde{S}_{ij}^{mb} = 2S_{ij}^{mb} - S_{ji}^{mb} .$$

$$\tilde{S}_{ej}^{ab} = 2S_{ej}^{ab} - S_{ej}^{ba}.$$

Ghh elements

$$u_m^i = \tilde{F}_k^b \tilde{S}_{ik}^{mb} - \hat{g}_{ib}^{kl} \tilde{S}_{kl}^{mb}.$$

$$\bar{\bar{g}}_j^i = \tilde{\tilde{F}}_j^i + u_m^i \delta_{mj}.$$

Gpp elements

$$u_{ae} = \tilde{F}_k^c \tilde{S}_{ek}^{ac} + \hat{g}_{cd}^{ld} \tilde{S}_{el}^{cd}.$$

$$\bar{\bar{g}}_{ac} = \tilde{\tilde{F}}_c^a + \delta_{ce} u_{ac}.$$

Gphph elements

$$\bar{\bar{g}}_{ab}^{ij}(J) = \tilde{J}_{ab}^{ij} + u_{ac}^{im} \delta_{mj} + u_{ae}^{ik} \delta_{eb} + u_{ae}^{im} \delta_{mj} \delta_{eb}.$$

where

$$u_{ac}^{im} = -\tilde{F}_k^c S_{ik}^{ma} + \hat{g}_{dc}^{la} \tilde{S}_{il}^{md} - \hat{g}_{cd}^{la} S_{il}^{md} + \hat{g}_{ic}^{kl} S_{kl}^{ma}.$$

$$u_{ae}^{ik} = \tilde{F}_k^c S_{ei}^{ac} + \hat{g}_{id}^{kl} \tilde{S}_{el}^{ad} - \hat{g}_{id}^{lk} S_{el}^{ad} + \hat{g}_{cd}^{ka} S_{ei}^{cd}.$$

$$u_{ae}^{im} = u_m^c S_{ei}^{ac} - u_e^k S_{ik}^{ma} + u_{il}^{mc} \tilde{S}_{el}^{ac} - u_{kd}^{im} S_{ek}^{ad} + \tilde{u}_{kd}^{im} S_{ek}^{da} + u_{ie}^{kl} S_{lk}^{ma}.$$

$$u_m^c = -g_{cd}^{kl} \tilde{S}_{kl}^{md}$$

$$u_k^e = g_{cd}^{kl} \tilde{S}_{el}^{cd}$$

$$u_{il}^{mc} = g_{cd}^{kl} \tilde{S}_{ik}^{md}$$

$$u_{kd}^{im} = g_{dc}^{kl} \tilde{S}_{mc}^{il}$$

$$\tilde{u}_{kd}^{im} = g_{dc}^{kl} S_{mc}^{il}$$

$$u_{ie}^{kl} = g_{cd}^{kl} S_{ei}^{cd}$$

Gphhp elements

$$\bar{\bar{g}}_{ab}^{ij}(K) = \tilde{K}_{ab}^{ij} + \tilde{\tilde{u}}_{ac}^{im}\delta_{mj} + \tilde{\tilde{u}}_{ae}^{ik}\delta_{eb} + \tilde{\tilde{u}}_{ae}^{im}\delta_{mj}\delta_{eb}$$

$$\tilde{\tilde{u}}_{ab}^{im} = \hat{g}_{ib}^{kl}S_{lk}^{ma} - \hat{g}_{ka}^{bc}S_{ki}^{mc} - \tilde{F}_{k}^{b}S_{ki}^{ma}$$

$$\tilde{\tilde{u}}_{ae}^{ij} = \bar{F}_{k}^{e}S_{ie}^{ac} + \hat{g}_{ja}^{cd}S_{ei}^{cd} - \hat{g}_{ic}^{kj}S_{ke}^{ac}$$

$$\tilde{\tilde{u}}_{ae}^{im} = u_{m}^{c}S_{ei}^{ca} - u_{e}^{k}S_{ki}^{ma} + \tilde{\tilde{u}}_{kd}^{im}S_{ke}^{ad} + \tilde{u}_{ie}^{kl}S_{lk}^{ma}$$

$$\tilde{\tilde{u}}_{kd}^{im} = g_{cd}^{kl}S_{li}^{mc}$$

$$\tilde{u}_{ie}^{kl} = g_{cd}^{kl}S_{ie}^{cd}$$

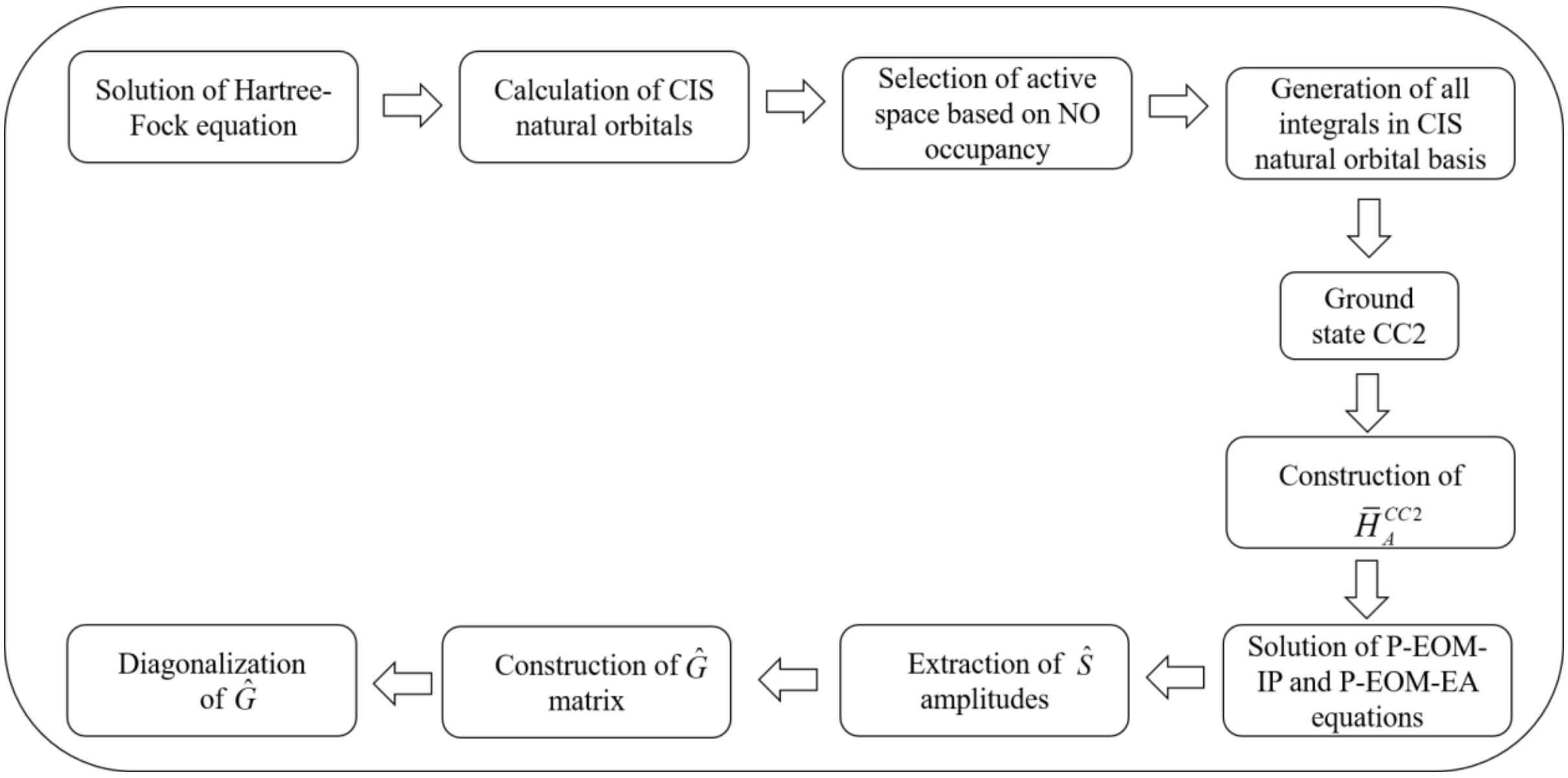

**Figure 1: Steps involved in ST-CC2 method**

**Table 1: The statistical analysis[a] of results for valence excited states (in eV).**

| | CC2<br>singlet/triplet | P-STEOM-MBPT2<br>singlet/triplet | ST-CC2<br>singlet/triplet |
|---|---|---|---|
| MAD | 0.27/0.56 | 0.503/0.462 | 0.376/0.396 |
| ME | 0.027/0.138 | 0.027/0.042 | 0.017/0.046 |
| AAD | 0.085/0.155 | 0.136/0.154 | 0.101/0.138 |
| RMSD | 0.105/0.199 | 0.167/0.195 | 0.125/0.175 |
| STD | 0.102/0.145 | 0.166/0.192 | 0.125/0.170 |

[a]The statistical quantities are abbreviated as maximum absolute deviation (MAD), mean error (ME), average absolute deviation (AAD), root-mean-square deviation (RMSD), and standard deviation (STD).

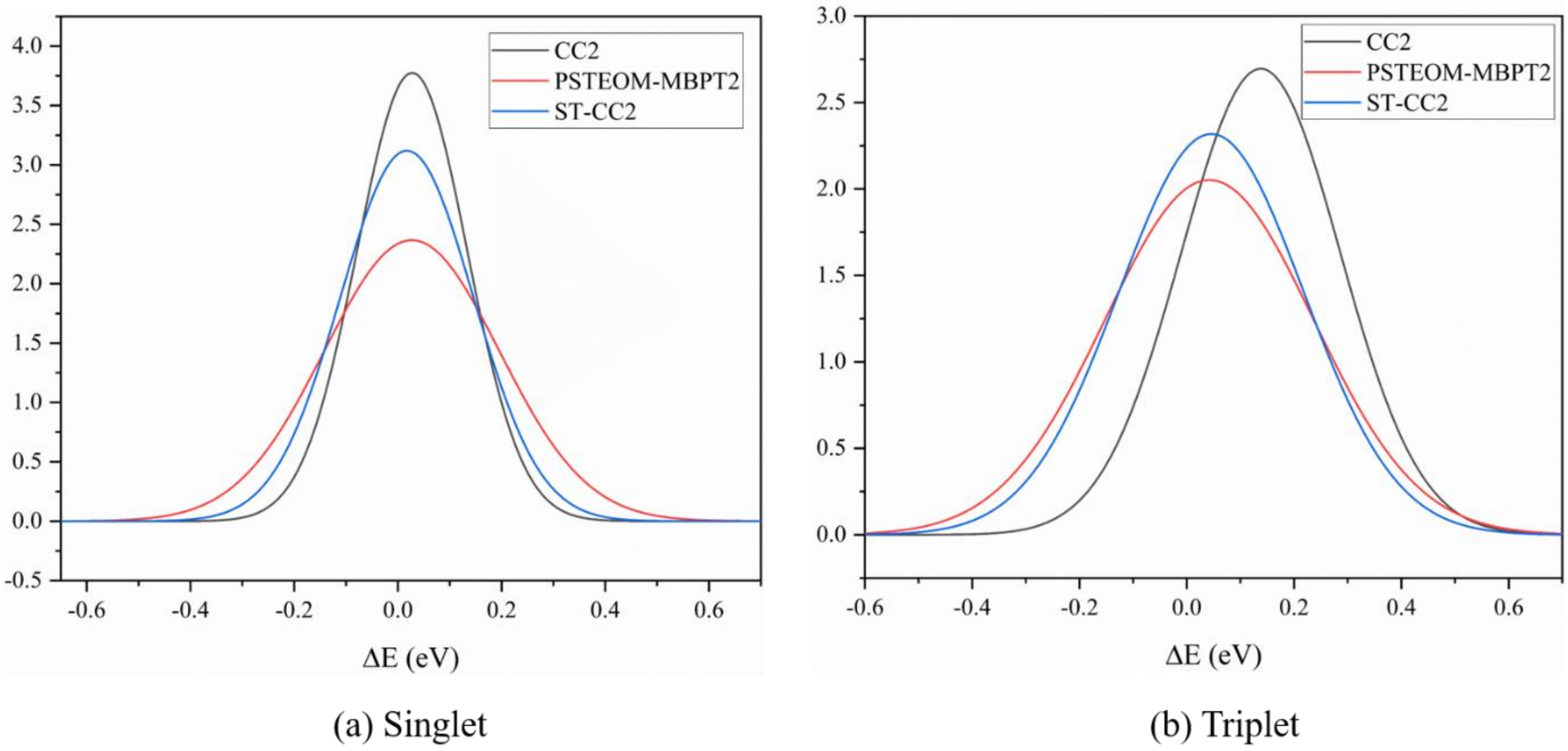

**Figure 2: Error distribution plot with respect to the CC3 reference values corresponding to the (a) singlet and (b) triplet valence excited states.**

**Table 2: The excitation energy of S-tetrazene in different approximate variants of STEOM-CCSD**

| State | Perecentsingles | CC3[a] | STEOM | ST-CC2 | Ext-ST-CC2 | P-STEOM-CCSD |
|---|---|---|---|---|---|---|
| $1^1B_{3u}$ | 89 | 2.53 | 2.482 | 2.663 | 2.808 | 2.282 |
| $1\ ^1A_u$ | 87.5 | 3.79 | 3.862 | 3.894 | 4.067 | 3.629 |
| $1^1B_{1g}$ | 84.6 | 4.98 | 5.011 | 5.212 | 5.293 | 4.866 |
| $1^1B_{2u}$ | 82.5 | 5.12 | 4.769 | 5.204 | 5.198 | 4.762 |
| $1^1B_{2g}$ | 80.7 | 5.34 | 5.429 | 5.622 | 5.747 | 5.253 |
| $2\ ^1A_u$ | 87.4 | 5.46 | 5.448 | 5.641 | 5.749 | 5.289 |
| $2^1B_{2g}$ | 79.2 | 6.23 | 6.52 | 6.528 | 6.698 | 6.282 |
| $2^1B_{3u}$ | 87.7 | 6.67 | 6.79 | 6.875 | 6.991 | 6.617 |
| $2^1B_{1g}$ | 84.7 | 6.87 | 6.956 | 6.987 | 7.167 | 6.721 |
| $3^1B_{1g}$ | 63.2 | 7.08 | 8.08 | 8.088 | 8.246 | 7.864 |
| $1^1B_{1u}$ | 91 | 7.45 | 7.578 | 7.629 | 7.882 | 7.299 |
| $2^1B_{1u}$ | 90.2 | 7.79 | 7.809 | 7.961 | 8.055 | 7.679 |
| $21B_{3g}$ | 63.6 | 8.47 | 9.029 | 8.996 | 9.155 | 8.836 |
| $2^1B_{2u}$ | 87.7 | 8.51 | 8.529 | 8.673 | 8.757 | 8.415 |

[a]The CC3 values are taken from reference 50

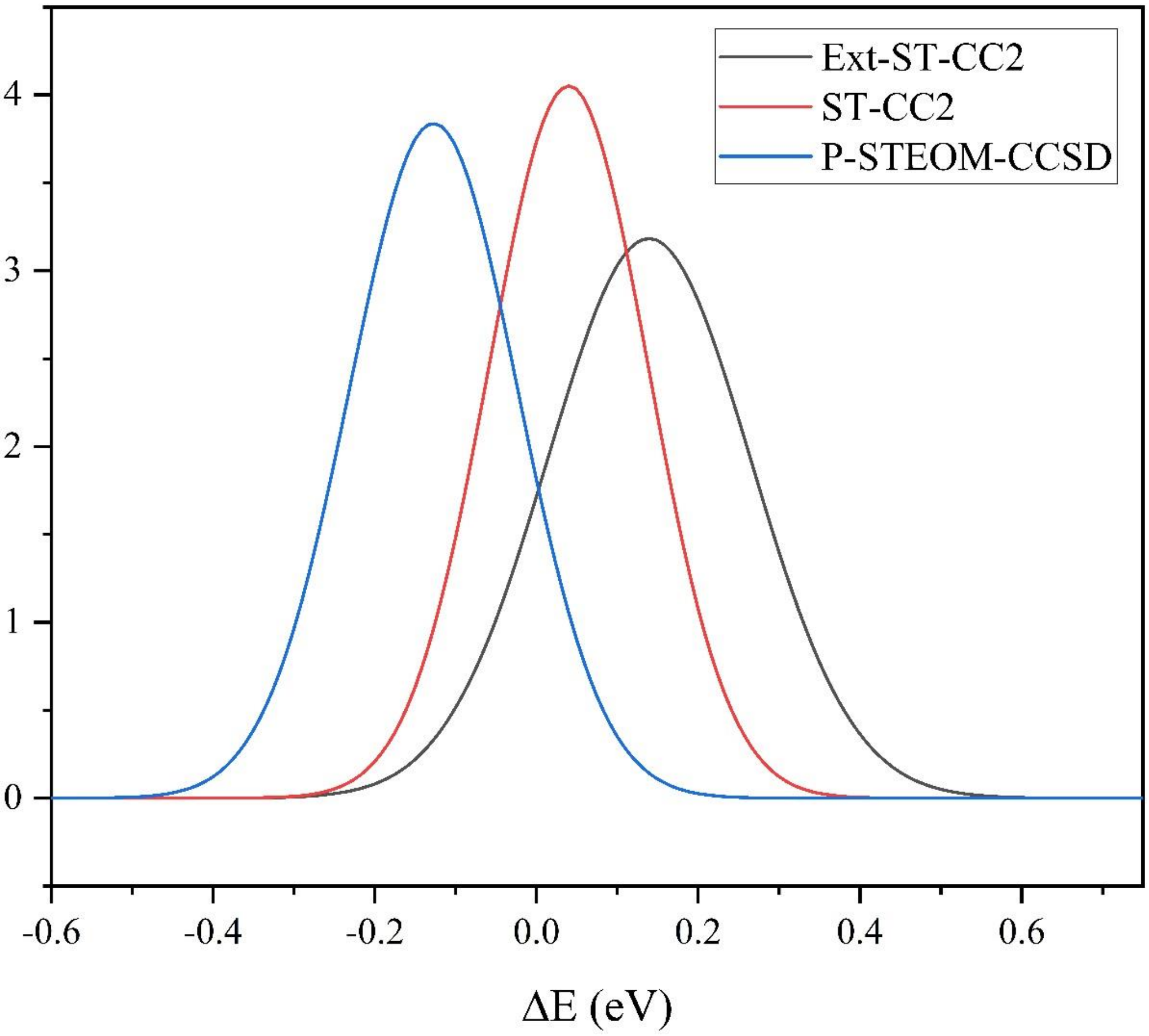

**Figure 3: Error distribution plot with respect to a STEOM-CCSD reference for valence singlet excited states of Thiel's test set.**

**Table 3: The statistical analysis[a] of results for Rydberg excited states (in eV)**

| | CC2<br>singlet/triplet | P-STEOM-MBPT2<br>singlet/triplet | ST-CC2<br>singlet/triplet |
|---|---|---|---|
| MAD | 0.985 | 0.375/0.449 | 0.276/0.309 |
| ME | -0.258 | 0.100/0.109 | 0.038/0.019 |
| AAD | 0.277 | 0.155/0.157 | 0.104/0.087 |
| RMSD | 0.373 | 0.174/0.176 | 0.126/0.112 |
| STD | 0.258 | 0.143/0.139 | 0.120/0.111 |

[a]The statistical quantities are abbreviated as maximum absolute deviation (MAD), mean error (ME), average absolute deviation (AAD), root-mean-square deviation (RMSD), and standard deviation (STD).

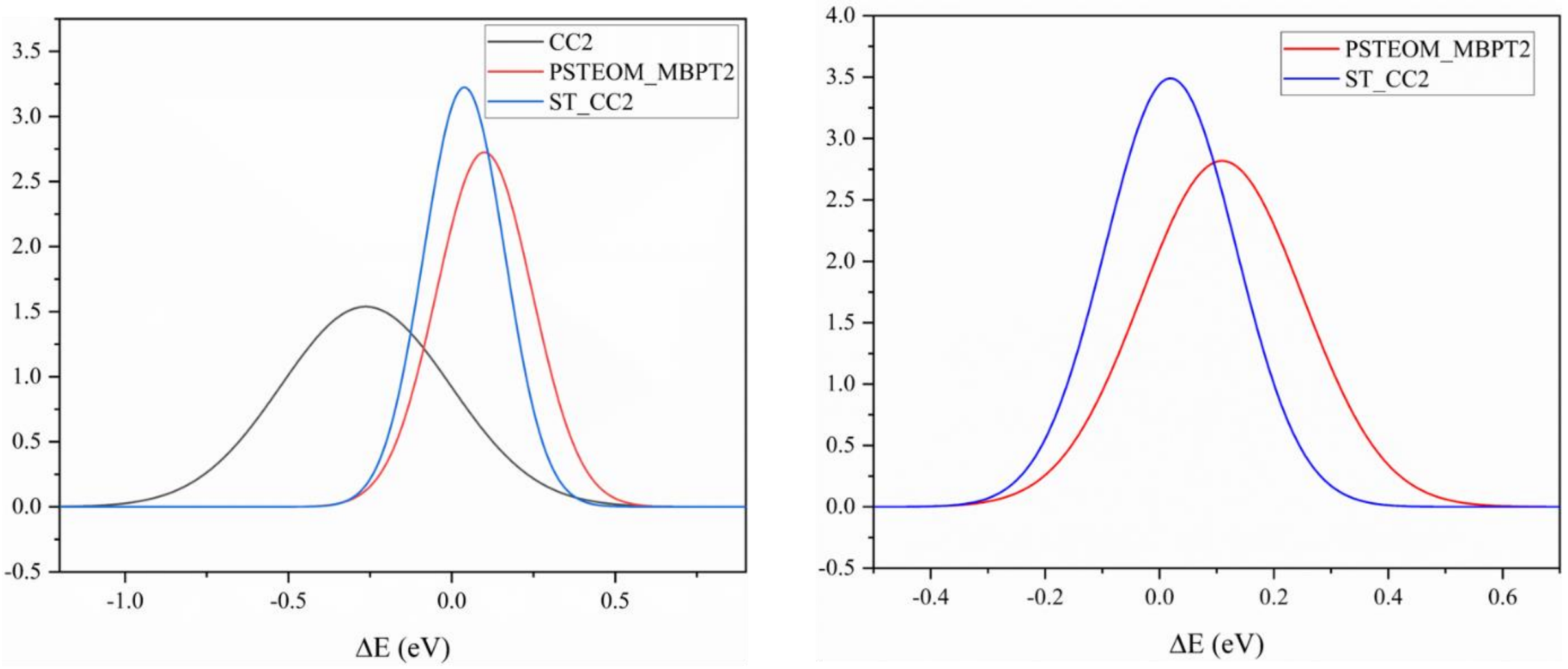

(a) Singlet (b) Triplet

**Figure 4: Error distribution plot corresponding to the (a) singlet and (b) triplet Rydberg excited states. CC3 and EOM-CCSD(T) method is used as a reference for the singlet and triplet states, respectively.**

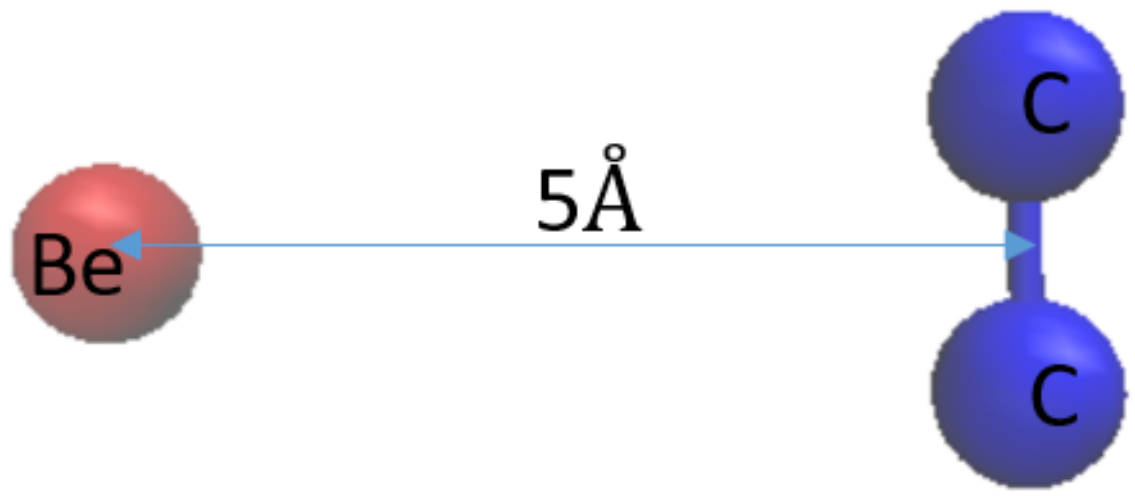

**Figure 5: Be–$C_2$ model complex, separated at 5 Å.**

**Table 4: Behavior of the second-order approximate coupled cluster methods in cc-pVDZ basis set for a charge transfer excitation in Be–$C_2$ model system with increasing distance between fragments (R)**

| [EE-(IP+EA)] (eV) | | | | |
|---|---|---|---|---|
| R (Å) | $-e^2R^{-1}$ (eV) | CC2 | P-STEOM-MBPT2 | ST-CC2 |
| 5 | -2.8800 | -2.4233 | -2.9371 | -2.9271 |
| 8 | -1.8000 | -1.4107 | -1.8015 | -1.8015 |
| 10 | -1.4400 | -1.0725 | -1.4325 | -1.4325 |
| 100 | -0.1440 | 0.1285 | -0.1440 | -0.1440 |
| 1000 | -0.0144 | 0.2490 | -0.0144 | -0.0144 |
| 10000 | -0.0014 | 0.2641 | -0.0014 | -0.0014 |

**Table 5: The statistical analysis of results for singlet charge transfer excited states (in eV)**

| | CC2 | P-STEOM-MBPT2 | ST-CC2 |
|---|---|---|---|
| MAD | 0.76 | 0.633 | 0.564 |
| ME | -0.363 | -0.041 | -0.026 |
| AAD | 0.363 | 0.187 | 0.163 |
| RMSD | 0.431 | 0.248 | 0.212 |
| STD | 0.240 | 0.254 | 0.218 |

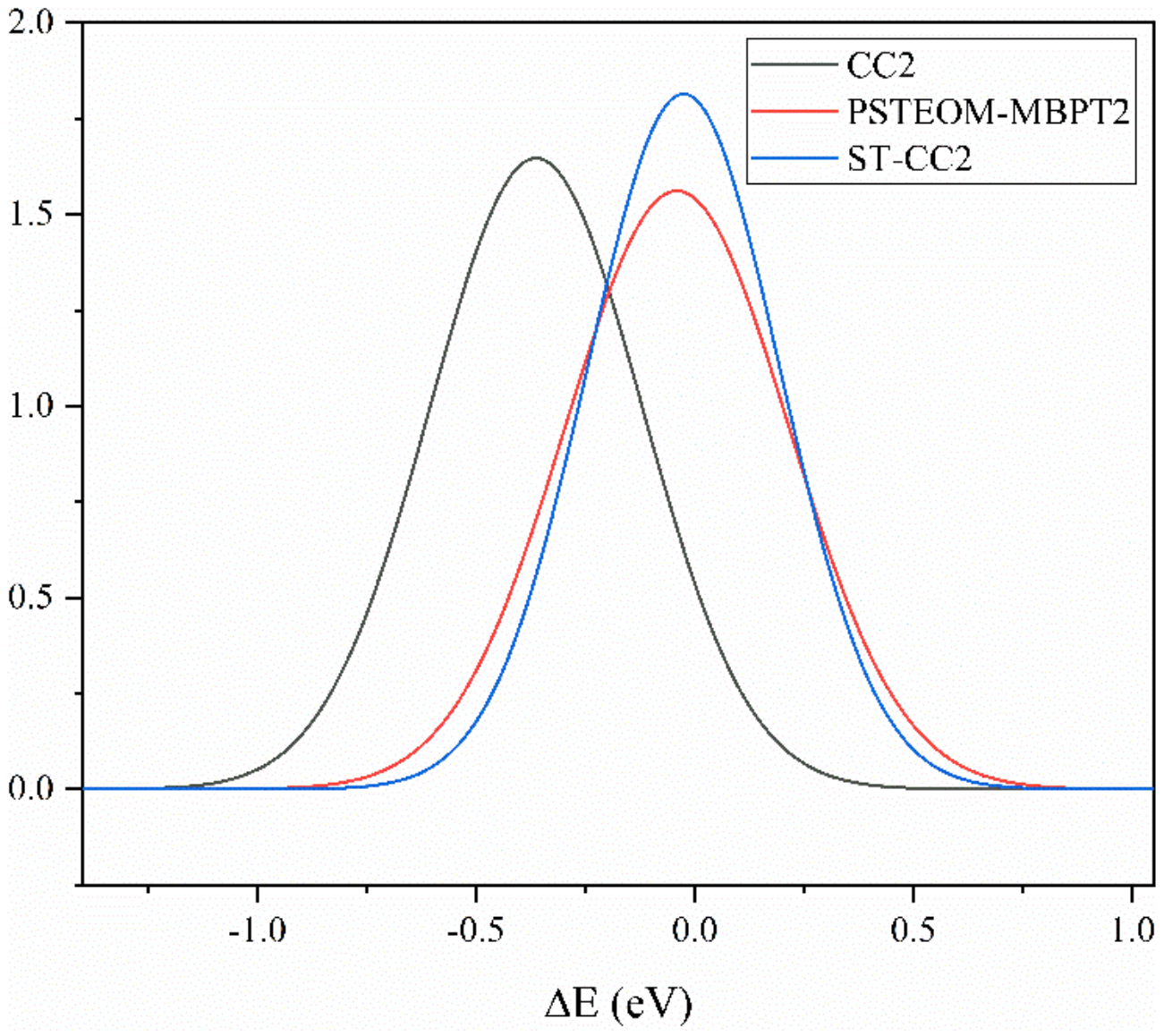

**Figure 6: Error distribution plot corresponding to the singlet CT excited states with respect to EOM-CCSDT-3 reference values**